# Oxygen molecular plasma at ambient temperature and elevated or high pressure

Yuri Kornyushin

*Maître Jean Brunschvig Research Unit, Chalet Shalva, Randogne, CH-3975*

Usually microscopic electrostatic field around ions is neglected when the ionization energy is concerned. The ionization energy is considered to be equal to that of a separate atom (molecule). Here the energy of the electrostatic field around ions is taken into account. It is shown that the energy of this field contributes to decrease in the effective ionization energy. The effective ionization energy may turn to zero at some critical concentration of delocalized electrons. This leads to a complete ionization of the atoms (molecules). Concrete calculations were performed for oxygen molecular gas.

## 1. Introduction

Regarding ionization problem it is assumed tacitly that the charge of the delocalized electrons is distributed uniformly throughout a sample [1]. The ionization energy is usually regarded to be equal to that of an isolated atom (molecule). It is well known that a uniform distribution of delocalized electrons in a sample is not equilibrium in the presence of the ions [1]. The charge redistributes itself, causing screening of the field of the ions [1]. This relaxation of the charge distribution leads to the decrease in the energy of a system. As a result the effective ionization energy decreases also.

When $N$ randomly distributed ions, charged with elementary charge $e$, and uniformly distributed charge of delocalized electrons are present, the electrostatic energy of a system consists of the energy of the ions [2,3], $e^2N/2r_0$ (here $r_0$ is some small distance, referring to the ion size), the energy of their charge distributed uniformly throughout a sample [3], the energy of the uniformly distributed charge of the delocalized electrons, and the energy of the interaction of the two uniform charges. The energy of the uniformly distributed ions charge, the energy of the uniformly distributed charge of the delocalized electrons, and the energy of their interaction, all three of them, annihilate together because the uniformly distributed negative and positive charges compensate each other. What's left is the energy of the bare ions, $e^2N/2\varepsilon r_0$.

When the delocalized electrons relax to equilibrium, screening the ions, the electrostatic energy of a sample decreases. This means that the effective ionization energy of the atoms (molecules) decreases also. Screening cuts off the long-range field, that is, leads to a decrease in the energy of electrostatic field.

## 2. Model

Let us consider charged with elementary positive charge $e$ ion in a gas. Usually microscopic electrostatic field around charged ions is neglected when the ionization energy is concerned. The ionization energy is considered to be equal to that of a separate atom (molecule) [1]. Recently it was shown that the microscopic electrostatic field contributes very essentially to thermodynamic properties [2]. The matter is that when a separate atom (molecule) is ionized, the electron becomes a plane wave in an infinite or very large space. This plane wave does not posses electric field and electrostatic energy. Before the ionization of the atom (molecule) we have electric fields of the charges of the ion and the localized electron. These fields contribute to the formation of the ground state energy of the atom (molecule). The field of the electron charge does not act on electron itself, only Coulombic electric field of the ion acts. But the energy of the electrostatic field of the charge of localized electron contributes to formation of the energy of a ground state. After the ionization of a separate atom (molecule) we have a Coulombic electrostatic field around the ion. Its energy contributes to the formation of a quantum ground state and its energy also as was mentioned. When we have many randomly distributed atoms (molecules) ionized, we have many delocalized electrons in a sample. The charge of these electrons redistributes itself to equilibrium distribution, causing screening of Coulombic field of the point charge of the ion and decrease in the electrostatic energy of a system.

We consider here the ions as point charges, and the delocalized electrons like a negatively charged gas.

The electrostatic field around a separate positive ion submerged, into the gas of delocalized electrons, is as follows [1]:

$$\varphi = (e/r)\exp{-gr}, \qquad (1)$$

where $r$ is the distance from the center of the ion and $1/g$ is the screening radius.

The electrostatic energy of this field is smaller than that of a bare ion. So the electrostatic energy of a system is also smaller as a result of the screening. The electrostatic energy of a separate ion with electrostatic field [Eq. (1)] is the integral over the volume of a sample of its gradient in square, divided by $8\pi$. The lower limit of the integral on $r$ should be taken as $r_0$, a very small value as was mentioned above. Otherwise the integral diverges. Taking into account that the value of the volume of a sample is usually very large (comparative to the ion volume) and performing calculation, we come to the following expression for the electrostatic energy of a separate ion [2]:

$$W = 0.5e^2(r_0^{-1} + 0.5g)\exp{-2gr_0}. \qquad (2)$$

As $gr_0$ is very small comparative to unity, Eq. (2) yields the following expression for the electrostatic energy of a separate ion [2]:

$$W = (e^2/2r_0) - 0.75e^2g. \qquad (3)$$

Here $e^2/2r_0$ is the electrostatic energy of the bare ion under consideration. When $g$ is small this term is the only one in the right-hand part of Eq. (3).

The decrease in the electrostatic energy of a separate ion due to the screening is $-0.75e^2g$. Electrostatic energy of $N$ randomly distributed ions is just $NW$ [3].

To ionize the first atom (molecule) in the presence of delocalized electrons the initial ionization energy $W_0$ is needed. Along with that the energy of a system decreases by $0.75e^2g/\varepsilon$, due to the relaxation, accompanied by screening. The effective ionization energy of the first ion is correspondingly as follows:

$$W_e = W_0 - 0.75e^2g. \qquad (4)$$

This problem was considered earlier in [4]. In [4] the ionization energy of impurities in semiconductors was discussed. The effective ionization energy was calculated in [4] for the case when the radius of a ground state is much smaller than the screening radius. The result was as follows:

$$W_e = W_0 - e^2g. \qquad (5)$$

This corresponds to the first term in the right-hand part of Eq. (2).

While calculating the effective ionization energy here no assumption, concerning the relation between the sizes of a ground state and the screening radius, was made.

Let us consider now a degenerate gas of delocalized electrons with concentration $n$. For this case the screening radius $1/g$ is Thomas-Fermi radius, defined by the following relation:

$$g^2 = 6\pi e^2 n/E_F, \qquad (6)$$

where $E_F = (\hbar^2/2m)(3\pi^2 n)^{2/3}$ is Fermi energy.

To calculate the effective ionization energy in this case, the kinetic energy of the delocalized electrons should be taken into account. The kinetic energy of the degenerate delocalized electrons per unit volume of a sample is as follows [1]:

$$T = 0.6 n E_F(n). \qquad (7)$$

The effective ionization energy per one molecule in the case regarded is accordingly as follows:

$$W_e = W_0 + (T/n) - 0.75 e^2 g. \qquad (8)$$

When the effective ionization energy is zero or negative, full ionization of the atom (molecule) occurs.

Let us perform further calculations for molecular gas of oxygen. In this case $W_e = 12.071$ eV [5]. Let us introduce another variable, $x$, according to the following relation:

$$n = 10^{24} x^6. \qquad (9)$$

Then the effective ionization energy $W_e$ is zero or negative when

$$x^4 \leq 1.3418 x - 0.5504. \qquad (10)$$

This happens when $0.43745 \leq x \leq 0.90076$, which corresponds to

$$7.0076 \times 10^{21} \text{ cm}^{-3} \leq n \leq 5.3414 \times 10^{23} \text{ cm}^{-3}. \qquad (11)$$

In the terms of the gas density, $d$, Eq. (11) yields $0.3727$ g/cm$^3 \leq d \leq 28.37$ g/cm$^3$.

It should be mentioned here that when total ionization of neutral gas occurs the delocalized electron concentration $n$ is equal to the number of the molecules per unit volume.

As at normal condition $n = n_0 = 2.687 \times 10^{19}$ cm$^{-3}$ [5], the range of the concentration discussed is from about 261 bar to 19.878 kbar or something in the terms of pressure.

### 3. Discussion

For gasses with high enough initial ionization energy $W_0$ the gas-plasma transition discussed here does not exist. The inequality, equivalent to Eq. (10), has no solution in this case accordingly. Oxygen molecular gas was chosen here because the ionization energy of the oxygen molecule is not too high. So there exists a rather wide range of the gas densities for which the gas-plasma transition may be realized.

Transition discussed here does not lead to the change in the number of molecules and their distribution. So from this part there is no change in the entropy of a system as a result of transition.

Delocalized electrons are degenerate. The range of Fermi temperatures is from 15,519.798 K to 281,523.1 K according to the calculations. That is we have a very strong degeneracy here. Every electron sits on its level and the contribution of the delocalized electrons to the change in the entropy of a system is negligible.

Possible influence of the dielectric constant of the gas was neglected also in the calculations performed. It should be noted, however, that the average distance between the molecules $1/n^{1/3}$ is larger from 3.02522 to 6.2293 times than the screening radius $1/g$ [see Eq. (6)]. So it looks like every molecule should be treated as a separate one without introducing dielectric constant into the calculations.

## References

1. C. Kittel, *Introduction to Solid State Physics*, Wiley, New York, 1996.
2. Yuri Kornyushin, arXiv:0710.3976v1.
3. G. F. Neumark, *Phys. Rev.* **B5**, 408, 1972.
4. Yu. V. Kornyushin, *Ukrainian Physical Journal*, **12**, 1327, 1967.
5. *CRC Handbook of Chemistry and Physics*, editor-in-chief David R. Lide, Boca Raton, Fla.: CRC Press, 2008.